\documentclass[conference]{IEEEtran}
\IEEEoverridecommandlockouts
\usepackage{cite}
\usepackage{amsmath,amssymb,amsfonts}
\usepackage{algorithmic}
\usepackage{graphicx}
\usepackage{textcomp}
\usepackage{xcolor}
\usepackage{subcaption}

\usepackage[printonlyused]{acronym} 
\usepackage[shortcuts,acronym, nomain]{glossaries}

\newacronym{lorawan}{LoRaWAN}{Long Range Wide Area Network}
\newacronym{lora}{LoRa}{Long Range}
\newacronym{physec}{PHYSEC}{Physical Layer Security}
\newacronym{iot}{IoT}{Internet of Things}
\newacronym{aes}{AES}{Advanced Encryption Standard}
\newacronym{cmac}{CMAC}{Cipher based Message Authentication Code}
\newacronym{dhke}{DHKE}{Diffie Hellman Key Exchange}
\newacronym{skg}{SKG}{Secret Key Generation}
\newacronym{mimo}{MIMO}{Multiple Input Multiple Output}
\newacronym{mmwave}{mmWave}{Millimeter-Wave}
\newacronym{lte}{LTE}{Long Term Evolution}
\newacronym{sram}{SRAM}{Static Random Access Memory}
\newacronym{kpi}{KPI}{Key Performance Indicator}
\newacronym{ip}{IP}{Internet Protocol}
\newacronym{crc}{CRC}{Cyclic Redundancy Check}
\newacronym{css}{CSS}{Chirp Spread Spectrum}
\newacronym{lpwan}{LPWAN}{Low Power Wide Area Network}
\newacronym{fsk}{FSK}{Frequency Shift Keying}
\newacronym{phy}{PHY}{Physical Layer}
\newacronym{mac}{MAC}{Media Access Control Layer}
\newacronym{mic}{MIC}{Message Integrity Code}
\newacronym{kgr}{KGR}{Key Generation Rate}
\newacronym{bdr}{BDR}{Bit Disagreement Rate}
\newacronym{rssi}{RSSI}{Received Signal Strength Indicator}
\newacronym{mqtt}{MQTT}{Message Queuing Telemetry Transport}
\newacronym{ccm}{CCM*}{Counter Cipher Block Chaining with Message Authentication Code}

\begin{document}

\title{Physical Layer Security based Key Management for LoRaWAN\\
\thanks{This is a preprint, the full paper is published in Proceedings of 2020 Workshop on Next Generation Networks and Applications (NGNA 2020). Personal use of this material is permitted. However, permission to use this material for any other purposes must be obtained from the authors. A part of this work was funded by the Federal Ministry of Education and Research (BMBF) of the Federal Republic of Germany (Foerderkennzeichen 16KIS1011, Industrial Radio Lab Germany). The authors alone are responsible for the content of the paper.}
}

\author{\IEEEauthorblockN{Andreas Weinand$^*$, Andreu G. de la Fuente$^{\dag}$, Christoph Lipps$^+$, Michael Karrenbauer$^*$}
\IEEEauthorblockA{$^*$Institute for Wireless Communication and Navigation (WICON), Technische Universität Kaiserslautern (TUK), Germany\\
$^\dag$Telecommunication School in Barcelona (ETSETB), Universitat Politècnica de Catalunya (UPC), Spain\\
$^+$Intelligent Networks Research Group, German Research Center for Artificial Intelligence, Kaiserslautern, Germany\\
Email: $^*$\{weinand, karrenbauer\}@eit.uni-kl.de, $^\dag$girones96@gmail.com, $^+$christoph.lipps@dfki.de
}
}

\maketitle

\begin{abstract}
Within this the work applicability of \gls{physec} based key management within \gls{lorawan} is proposed and evaluated using an experimental testbed. Since \gls{iot} technologies have been arising in past years, they have as well attracted attention for possible cyber attacks. While \gls{lorawan} already provides many of the features needed in order to ensure security goals such as data confidentiality and integrity, it lacks in measures such as secure key management and distribution schemes. Since conventional solutions are not feasible here, e.g. due to constraints on payload size and power consumption, we propose the usage of \gls{physec} based session key management, which can provide the respective measures in a more lightweight way. The results derived from our testbed show that it can be a promising alternative approach.
\end{abstract}


\section{Introduction}
\label{intro}
Since the upcoming of \gls{iot} applications, there have been many radio technologies proposed as enablers for the transmission of data from end devices, such as sensor nodes, towards cloud or other central processing entities. These can provide advantages in the sense of a higher deployment flexibility and enable the connection of a huge number of devices at a low cost, compared to wired systems. On the other hand, they bring challenges and risks due to the open nature of the wireless channel. Especially in industrial scenarios, such as e.g. smart metering, agricultural applications or process monitoring, the nondisclosure of intellectual property such as process control parameters, machine configuration data or even simple information such as the production volume have to be ensured. Beside online attacks interfering with such applications and causing damage instantly, other risks such as blackmailing have increased recently as well.\\
In order to prevent such cyber attacks, e.g. symmetric key cryptography ciphers such as the \gls{aes} \cite{NIST2001} can be used to ensure data confidentiality and integrity. Both of these requirements are fulfilled by the \gls{lorawan} \cite{lora_spec} protocol, which utilizes the AES-128 cipher suite for data encryption and decryption and \gls{aes} based \gls{cmac} \cite{IETF2006} as message integrity code. Since the keys used for the respective \gls{aes} operation are typically derived manually from device manufacturers, this offers a high possibility for disclosure. Additionally, the root key is typically hard coded on both sides, end device and the \gls{lorawan} network or application server. This brings the problem, that it can not be refreshed regularly, in order to enable security concepts such as perfect forward secrecy. Conventional key management schemes, such as e.g. \gls{dhke}, are not applicable here due to their high requirements towards computational power and transmission overhead. Further, the key management should be realized at a high level of usability, where no manual configuration is required by e.g. a system administrator. This is especially due to scalability reasons, occuring e.g. in massive \gls{iot} scenarios. All these requirements can be fulfilled by \gls{physec} based key generation, where the idea is to exploit the characteristics of the wireless channel as a random process and derive a secret key from that. There are however some other conditions to be fulfilled, such as channel reciprocity between two parties deriving a secret key. That means, the time and frequency at which both of them sample the channel have to be aligned. Resulting from that, key bits derived from that process might not be identical between two parties and require further processing and communication for information reconciliation. Therefore, it is desired to keep the erroneous bits before that stage as low as possible by applying optimal quantization and reciprocity enhancement schemes. Further, it has to be ensured, that initial trust is set up between involved parties, such as a cryptographic authentication process. A periodic session key refreshment procedure can then be realized by support of \gls{physec} based \gls{skg} on top of that trust root.\\
The remaining work is structured as follows, within section \ref{related_work} we present related work considering \gls{physec} and especially the concept of \gls{skg}. In section \ref{lorawan} we introduce the \gls{lorawan} protocol and within section \ref{physec_lorawan}, the \gls{physec} based key generation procedure is presented. Section \ref{evaluation} elaborates the results derived from our testbed and section \ref{conclusion} concludes our work.

\section{Related Work}
\label{related_work}
Previous works have already proven, that the wireless channel can be used as a good source of randomness in order to generate symmetric key pairs. E.g. the application of \gls{physec} based key generation was already investigated for IEEE 802.11 systems in \cite{Zhang2016}, where a testbed was developed using the Wireless open-Access Research Platform (WARP) and different radio features are evaluated within different scenarios for the purpose of \gls{skg}. In \cite{Yang2015}, the usage of \gls{physec} in massive \gls{mimo} systems and \gls{mmwave} communication in heterogeneous networks is investigated. Further, works have also investigated cellular systems for application of \gls{physec}, such as \cite{Lipps2019} or \cite{Wang2018}, where Device-to-Device (D2D) communication in \gls{lte} networks with an underlying cellular network infrastructure is used. Another work studies the application of \gls{physec} within the downlink of cellular networks by considering different serving base stations scenarios \cite{Geraci2014}. Among the first works considering the specific demands of securing \gls{iot} applications based on \gls{physec} were e.g. \cite{Guillaume2015}, \cite{Ambekar2012} and \cite{Zenger2014}. The authors in \cite{Zenger2015a} are especially considering resource constrained end devices and give an experimental proof, that the energy consumption of \gls{physec} based key management can be decreased by more than one order of magnitude compared to conventional approaches (e.g. \gls{dhke}). Some works have also tried solving both, authentication and key generation based on \gls{physec} in terms of combining Physical Unclonable Functions (PUFs) derive from \gls{sram} memory characteristics \cite{Lipps2018} with the \gls{physec} key generation approach \cite{Lipps2020}.
In \cite{AdefemiAlimi2020} a security analysis of the \gls{lorawan} protocol is provided. Other works have therefore proposed the usage of \gls{physec} within \gls{lorawan}. E.g. \cite{Robyns2017} proposes the use of unique \gls{lora} chipset characteristics as method for radio fingerprint identification. Other works propose the use of channel characteristics for \gls{physec} based key generation in different scenarios such as \cite{Ruotsalainen2020}, \cite{Zhang2018}, \cite{Xu2019a}. These works provide a good investigation of the actual radio characteristics and maximum achievable \gls{bdr} but do not provide deep insights towards performance in terms of the \gls{kgr}. Therefore, we put our focus here on the latter and evaluate the key generation scheme based on that within typical \gls{iot} scenarios. Further, we strictly follow the \gls{lorawan} protocol, which e.g. \cite{Xu2019a} does not follow.


\begin{figure}[t]
	\begin{center}
		\includegraphics[width=\columnwidth]{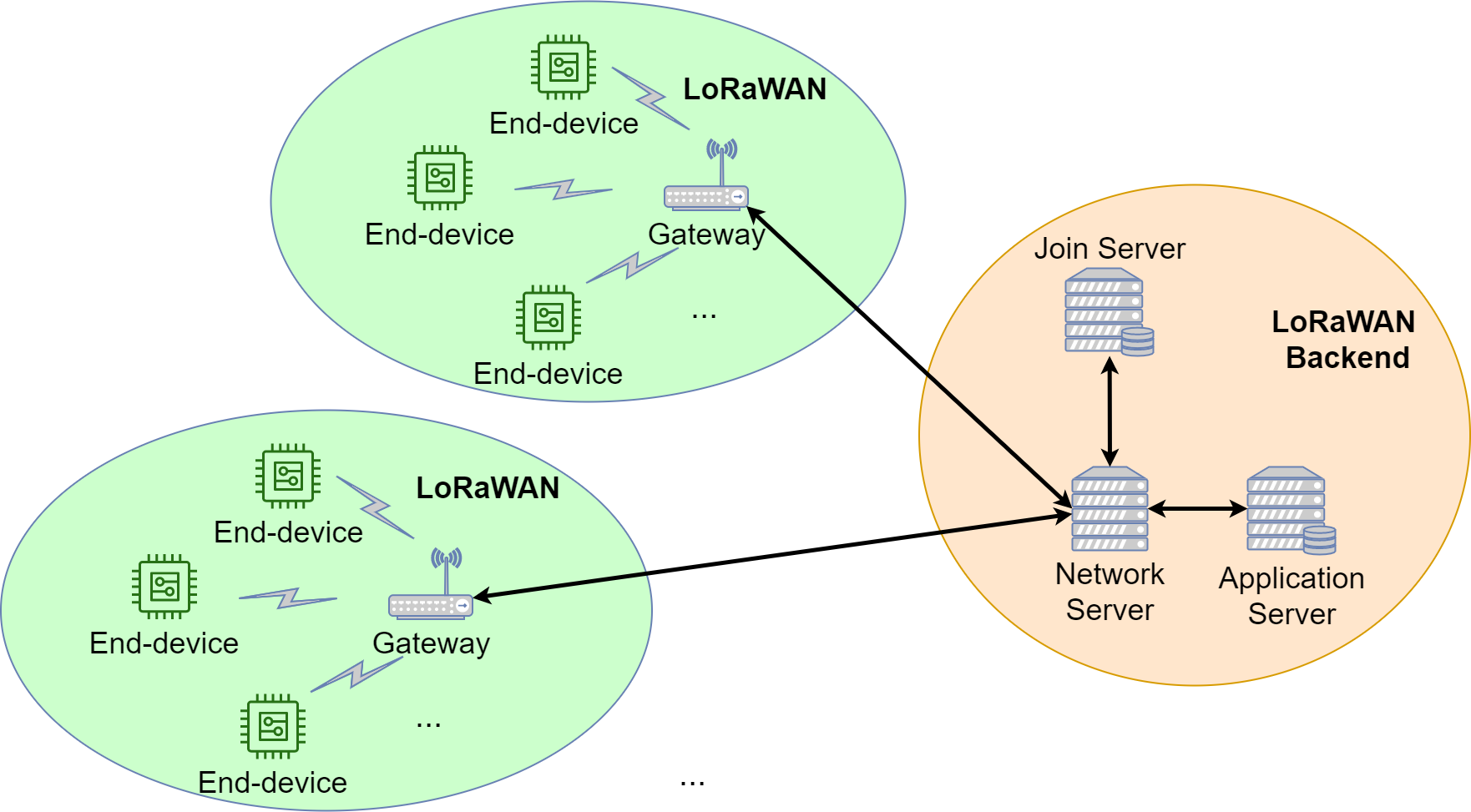}
		\caption{\gls{lorawan} system architecture} 
		\label{lora_arch}
	\end{center}
\end{figure}
\section{\gls{lorawan} Protocol}
\label{lorawan}
\gls{lorawan} denotes the high layer protocol for \gls{lora} based radio transceivers, whereas \gls{lora} itself is a physical layer protocol patented by Semtech \cite{Seller2016}. The architecture of a typical \gls{lorawan} system is given in Fig. \ref{lora_arch} and is usually deployed as a star-of-stars topology since many end devices are connected to a gateway and many gateways on the other side are connected towards a network server. The network server has the task to manage the configuration of radio links towards all end devices and route all the packets received from the gateways to the corresponding application server respectively. It further manages the downlink transmissions from the application server towards the device side. The task of the gateway is to translate between the \gls{ip} based communication towards the network and application servers on the backend side and the \gls{lorawan} protocol towards the device side. Another entity is the join server, which is in charge of managing the cryptographic root keys of the users and deriving session keys from these. The application server however receives all the packets that the network server routes to it and there can exist multiple application servers and applications served by a single \gls{lorawan} network, since the application payload is secured within each application, independent of the network server operations.


\subsection{PHY \& MAC Layer}


\gls{lora} uses a spread spectrum modulation which is based on \gls{css}. \gls{css} is a low power consumption modulation that uses up and down chirps for data transmission. It is robust against interferences such as multipath fading and doppler effects due to its high Bandwidth-Time product. Since it is typically deployed within the Industrial Scientific and Medical (ISM) bands, it does not suffer from inter-system interference as much as other \gls{lpwan} systems (e.g. IEEE 802.15.4). Different bandwidth options ranging from 125 kHz to 500 kHz are provided. Due to regulatory aspects such as duty cycle restrictions, typically only a bandwidth of 125 kHz is used (Europe), which also decreases the power consumption within the end devices. Further, \gls{lora} supports the encoding of data symbols using either 6 different spreading sequences or \gls{fsk} modulation. For the coding options using spreading sequences, data rates within the range from 0.3 kbps to 11 kbps can be enabled, wheras for \gls{fsk} modulation the data rate is always 50 kbps. Due to the orthogonality between the spreading sequences of different spreading factors (ranging from $2^7$ to $2^{12}$), they can be used for code multiple access at the same time and frequency resources. Further, this allows for adjusting the capacity, transmission reliability and communication range.

\begin{figure}[t]
	\begin{center}
		\includegraphics[width=\columnwidth]{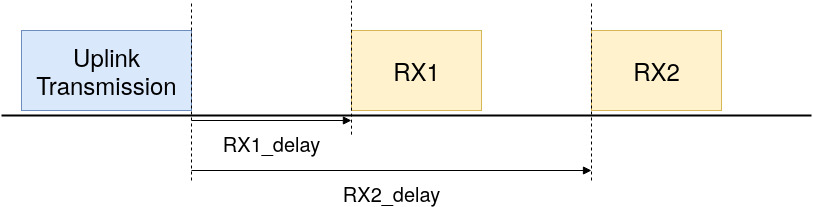}
		\caption{\gls{lorawan} class A receive window options} 
		\label{lora_receive_window}
	\end{center}
\end{figure}
The \gls{mac} includes features such as \gls{phy} configuration (called \gls{mac} Commands), and management of the medium access strategies, which depend on the class type of the end device. The \gls{phy} configuration typically includes the setting of used ISM band, used bandwidth, transmission power and spreading factor. In newer released (V1.1) there is also a feature added in order to optimize the radio access utilization, called Adaptative Data Rate (ADR), selecting the adequate data rate (spreading factor) and frequency band for the current interference situation between the end device and gateway respectively.\\
There are three different end device classes existing in \mbox{\gls{lorawan}}, which are class A, class B and class C devices. The main difference between the classes is the designated time to listen for a received message from the gateway by opening their receive windows respectively. Therefore, each class performs at different power consumption levels. Class A is defined as baseline implementation that must be supported by all \gls{lorawan} end devices. For uplink channel access, class A devices deploy the ALOHA protocol. In the downlink however, receive windows are only used for the reception of acknowledgement messages from the gateway following an successful uplink transmission. The end device offers two time slots during which it activates its receive chains as depicted in Fig. \ref{lora_receive_window}. Otherwise they are in idle mode and no other receive windows are allowed. The receive windows are opened after a delay time depending on the legislation of the territory where it is used. E.g. according to the region parameters specification \cite{lora_regional}, within the EU868 band, the default values are 1 and 2 seconds for the first and second receive windows respectively. Beside power consumption limitations of the end devices, another reason for that strategy is the duty cycle limitation for downlink transmissions in case of highly scaled scenarios, such as massive \gls{iot}. Class B and C are extensions of class A and offer more options for receive windows by implementing a periodic (class B) or permanent (class C) strategy for that.

\begin{figure*}[ht!]
	\begin{center}
		\includegraphics[width=\textwidth]{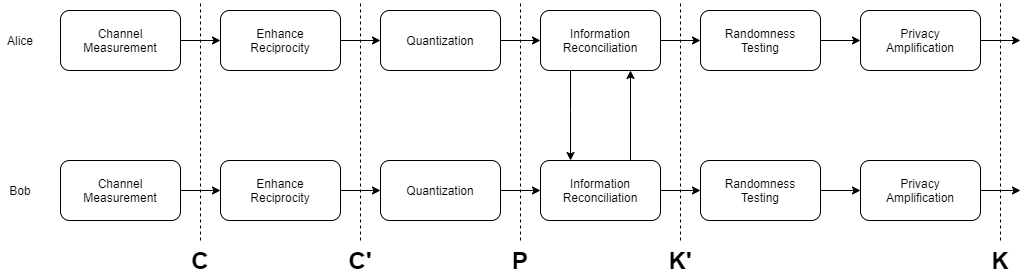}
		\caption{Procedure of \gls{physec} based key generation} 
		\label{physec_skg}
	\end{center}
\end{figure*}
\subsection{Security}
Within \gls{lorawan}, data encryption and \gls{mic} operations are supported. In order for parties to identify and authenticate mutually, unique identification by EUI-64 (IEEE 64-Bit Extended Unique Identifier) addresses is used. This enables the device side to uniquely identify and authenticate towards the backend side using their respective device EUI (DevEUI), whereas backend devices, such as the join server (JoinEUI), can identifiy themself towards the end devices, e.g. during the join procedure. For the join procedure within a \gls{lorawan} network, two methods are supported. Either Over the Air Activation (OTAA), where a join procedure is executed or by Activation by Personalization (ABP), where session keys are installed manually. Therefore, the respective EUI has to be stored in an end device before initiating a join procedure (join request) in case of OTAA.

\subsubsection{Key Management}
Due to the power constraints of typical \gls{lorawan} applications presented in section \ref{intro}, the protocol provides symmetric key cryptography in order to secure transmissions between end devices and gateways, whereas the communication between the gateways and other system components (e.g. network server) is secured by standard \gls{ip} based solutions such as e.g. Transport Layer Security (TLS). There are two root keys, which are the used for derivation of further session keys, which are the Application Key (AppKey) and Network Key (NwkKey) in the protocol version 1.1. The different keys are used for message transmission from the end device towards the different destinations. If a message is addressed towards the application server, the Application Session Key (AppSKey) is used for encryption and decryption, whereas for communication towards the network server (e.g. \gls{mac} commands), the Network Session Key (NwkSEncKey) is used. As mentioned before, two methods for device activation are available in \gls{lorawan} (OTAA and ABP) in order to install the session keys. The ABP option has the drawback, that the same session keys are used throughout the device lifetime (or until ABP is executed again), whereas in OTAA network session keys are derived during the join procedure based on the NwkKey. Therefore, \cite{lora_spec} recommends to use OTAA for a higher level of security. Further, the separation of keys for network server and application server traffic allows the usage of federated network servers, since application traffic confidentiality is still guaranteed. Additionally, OTAA allows for roaming scenarios, where end devices can join the networks of other providers and receive respective network session keys, as network session keys can be changed by the network server.
On the device side it is recommended to store the root keys (AppKey and NwkKey) in a way, such that the extraction of the keys and reusing by malicious actors is prevented. On the backend side however, the storage of root keys and associated key derivation operations for session keys is realized by terms of the Join Server. \gls{lorawan} delegates the responsibility of maintaining the root keys secretly to the user. If the root keys are revealed the system is totally vulnerable. Therefore, it is desirable to not keep the same root key throughout the whole device lifecycle and derive session keys only there.

\subsubsection{Security operations}
The respective session keys are used in order to protect the data confidentiality and integrity at the \gls{mac} layer of the \gls{lorawan} protocol. For both, the AES-128 cipher suite is used. For a default uplink transmission towards the application server, the \gls{mac} payload is first encrypted using the AppSKey in enhanced \gls{ccm} mode. For downlink messages however, the network server uses the respective AES decryption method for ciphering, which allows the end device to decipher this encrypted message using the encryption operation. Due to that, end devices to not need to have the AES decryption operation implemented, reducing cost and complexity. 
After payload encryption, the \gls{mic} is calculated over the \gls{mac} frame header, port field and \gls{mac} payload by using the AES-CMAC algorithm. In \gls{lorawan}, the integrity is protected in a hop-by-hop fashion. That means, that transmissions between the end device and the network server and transmissions between the network server and the application server have independent integrity protection. This however makes \gls{lorawan} vulnerable to malicious network servers, e.g. by making unauthorized changes to the header metadata. Additionally to encryption and \gls{mic} operation, counters are used for both directions (uplink and downlink) in order to prevent replay attacks. These counters are fields within the \gls{mac} header and therefore protected by the \gls{mic}. Since they have only a length of 8 Bit, replay attacks can not be completely prevented.

\section{\gls{physec} based key management}
\label{physec_lorawan}

\begin{figure}[b]
	\begin{center}
		\includegraphics[width=0.75\columnwidth]{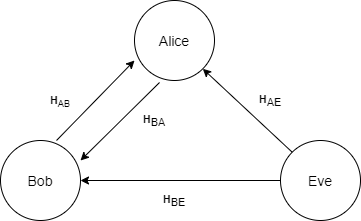}
		\caption{\gls{physec} system model} 
		\label{physec_system_model}
	\end{center}
\end{figure}

In \gls{physec} based key management, the goal is to derive a secure session key from the wireless channel characteristics that can be used between two parties, e.g. Alice and Bob. We denote the channel characteristics, measured by either Alice or Bob $\mathbf{H}_{AB}$ and $\mathbf{H}_{BA}$ respectively. Additionally, an adversary Eve can be present and eavesdrop the ongoing traffic. In such a case, $\mathbf{H}_{EA}$ are the channel characteristics measured by Eve in case of a transmission by Alice and $\mathbf{H}_{EB}$ in case of a transmission by Bob. The \gls{physec} system model is shown in Fig. \ref{physec_system_model}. The principle of channel reciprocity indicates that the channel measurements $\mathbf{H}_{AB}$ and $\mathbf{H}_{BA}$ are equal when they are conducted during the coherence time of the channel and in the absence of noise. Further, the channel decorrelation property denotes, that the characteristics change based on the current transmitter and receiver locations. It can be assumed, that the channel between a pair of transceivers is already decorrelated, when one of them changes their location by more than $\frac{\lambda}{2}$. Therefore, it has to be guaranteed, that Eve is not located closer as $\frac{\lambda}{2}$ towards Alice ($d_{EA} > \frac{\lambda}{2}$) or Bob ($d_{EB} > \frac{\lambda}{2}$), where $\lambda$ is the wavelength of the transmitted signal and $d_{EA}$ and $d_{EB}$ the distance between Eve and Alice
and Bob respectively. E.g. if the EU868 frequency bands are considered, the wavelength corresponds to $\lambda\approx 34.56$ cm. This can e.g. be achieved by physically restricted access in rooms where end device nodes or gateways are deployed.
In order to negotiate a secret key of length $L = MN$ Bits only known to Alice and Bob, several steps need to be executed as shown in Fig. \ref{physec_skg}. First, both parties Alice and Bob need to measure their channel mutually over a period of M measurements, e.g. by means of channel probing. This will yield to channel profiles $\mathbf{C} = (\mathbf{H}^{(1)}, \mathbf{H}^{(2)}, \ldots, \mathbf{H}^{(M)})$. To enhance the reciprocity of the channel (unmatching values in the channel measurements due to e.g. noise), different approaches, such as e.g. kalman filtering \cite{Ambekar2012} or polynomial curve fitting, have been considered. In the next step, the enhanced channel profiles $\mathbf{C'}_{AB}$ and $\mathbf{C'}_{AB}$  are derived by both, Alice and Bob, are quantized in order to obtain preliminary keys $\mathbf{P} = (P^{(1)}, P^{(2)}, \ldots, P^{(L)})$. In general, N Bits of the preliminary key are derived from each channel measurement $\mathbf{H}^{(k)} (k=1, \ldots, M)$ by applying a respective quantization scheme. Due to remaining unmatched values in the enhanced channel measurements, a disagreement of Bits in preliminary keys can still exist ($\mathbf{P}_{AB} \ne \mathbf{P}_{BA}$). These errors are detected and corrected in the information reconciliation stage by means of error correction coding, e.g. turbo codes or Bose–Chaudhuri–Hocquenghem (BCH) codes, yielding the synchronized key $\mathbf{K'} = \mathbf{K'}_{AB} = \mathbf{K'}_{BA}$. Previous works could proof, that a \gls{bdr} of up to $20\%$ is still tolerable for some codes in order to recover the key bits. Due to parity information exchange between Alice and Bob to match their keys during the information reconciliation stage, an attacker can use these information in order to gain partial knowledge of the key. To reduce this effect, as well as to enhance the entropy of the key, privacy amplification is utilized after testing the key randomness. A common approach here are cryptographic hashing algorithms such as the Secure Hashing Algorithm (SHA), e.g. SHA-2, SHA-3. Finally Alice and Bob both share the secret key $\mathbf{K} = \mathbf{K}_{AB} = \mathbf{K}_{BA}$. The performance of \gls{physec} based key generation is typically measured by two metrics. The \gls{kgr} which is the effective rate of generated bits of the symmetric key per second. The other one is the \gls{bdr}, which denotes the amount of disagreeing bits between two parties before the information reconciliation stage. Since within the present scope, the frequency of messages being transmitted between Alice and Bob is quite low, the \gls{kgr} is more relevant compared to the \gls{bdr} and therefore especially evaluated in section \ref{evaluation} within our testbed.

\begin{figure}[t!]
\centering
\begin{subfigure}[b]{\columnwidth}
   \includegraphics[width=1\linewidth]{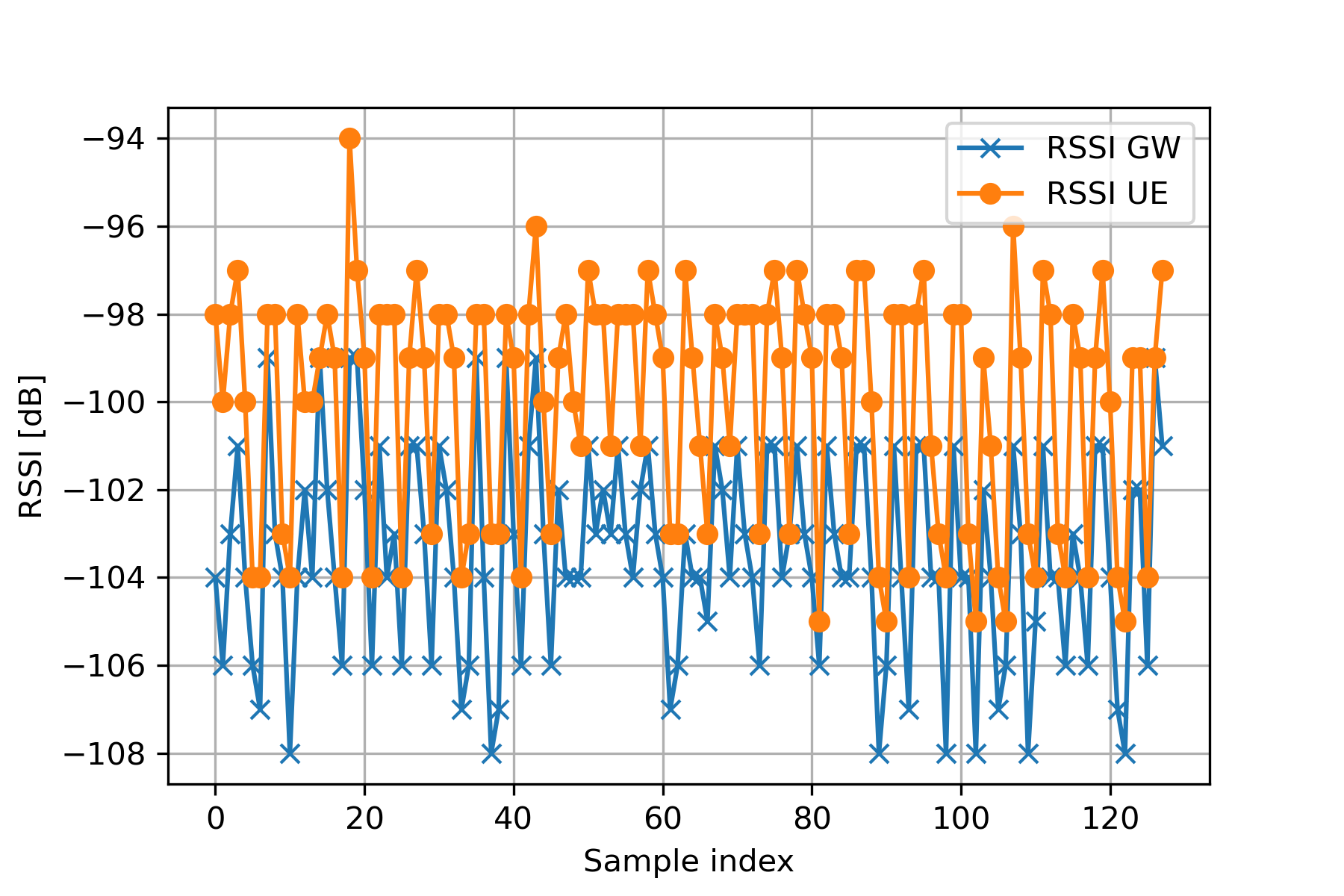}
   \caption{}
   \label{rssi_absolute} 
\end{subfigure}

\begin{subfigure}[b]{\columnwidth}
   \includegraphics[width=1\linewidth]{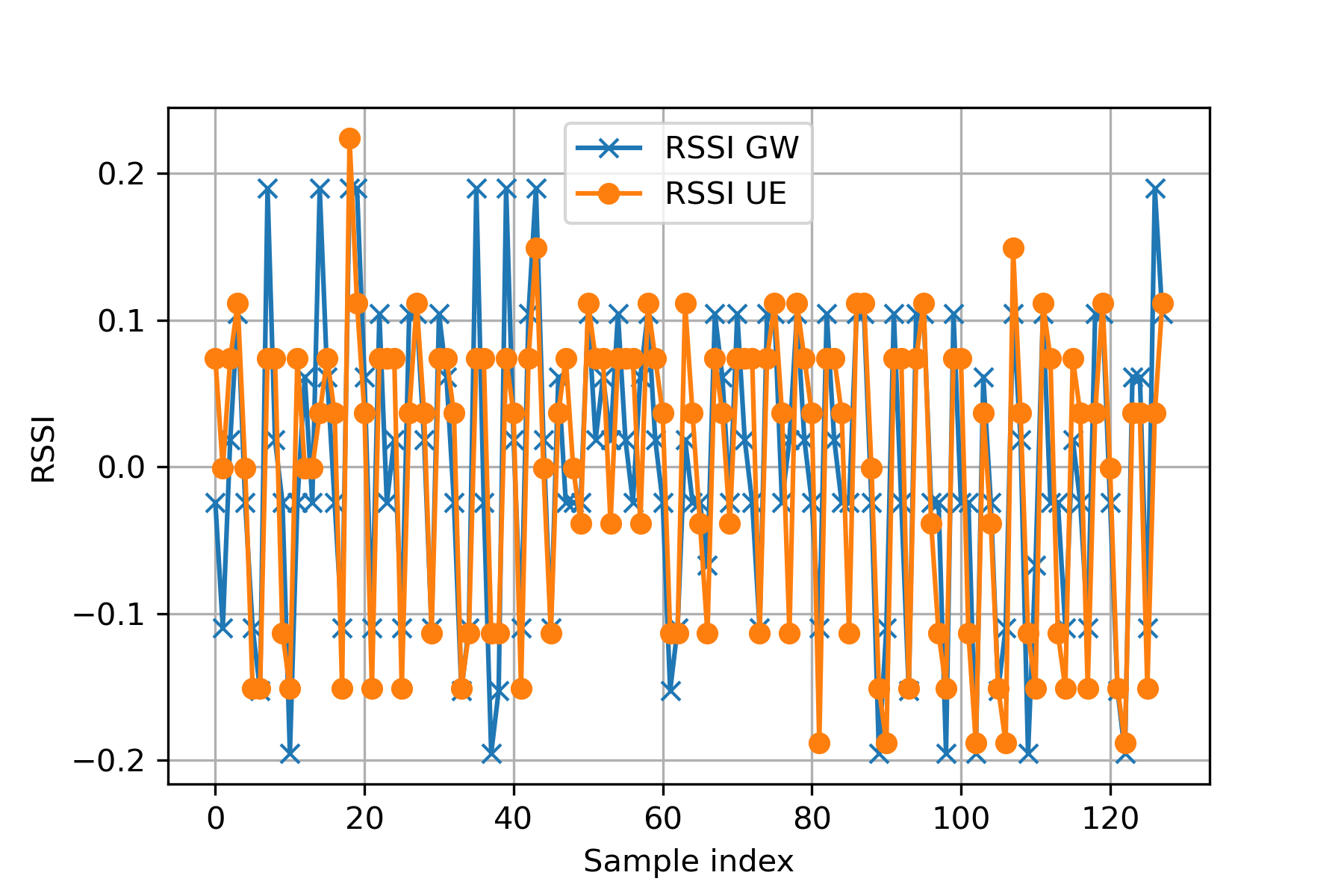}
   \caption{}
   \label{rssi_scaled}
\end{subfigure}

\caption[]{Exemplary \gls{rssi} capture of size 128 samples from Alice and Bob (a) in absolute dB (b) scaled by mean and unit variance}
\label{rssi_captures}
\end{figure}
\section{Evaluation}
\label{evaluation}
Within this section, we first present the our \gls{lorawan} based testbed. Then we show the results derived from the acquired dataset and discuss them.

\begin{figure}[b!]
	\begin{center}
		\includegraphics[width=\columnwidth]{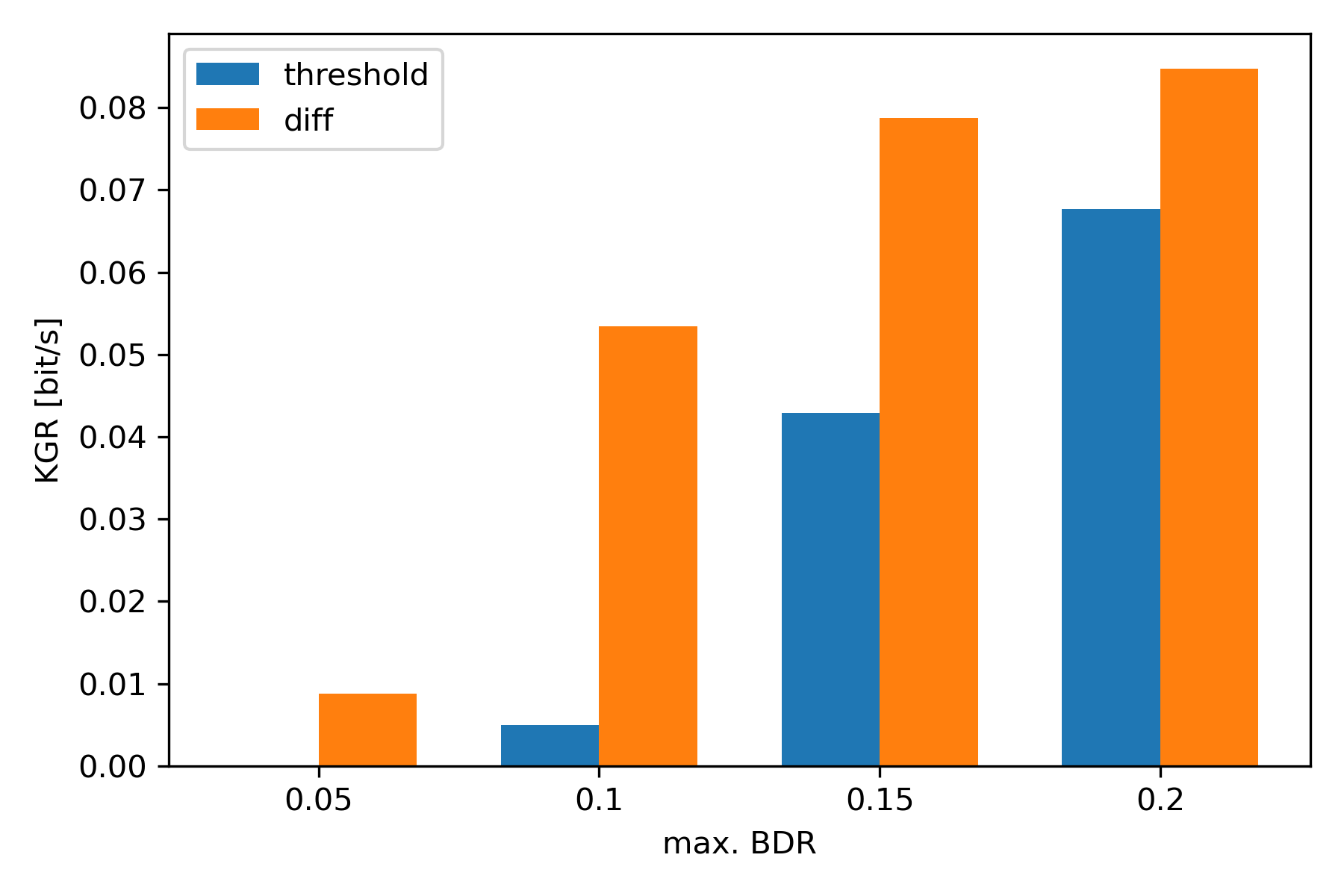}
		\caption{\gls{kgr} for maximum allowed preliminary key \gls{bdr} and quantization method} 
		\label{kgr_dep_bdr}
	\end{center}
\end{figure}
\subsection{Experimental Testbed}
In order to record channel state parameters, such as the \gls{rssi}, Signal-to-Noise Ratio (SNR), we use the Adeunis \gls{lorawan} fieldtest device in class A device configuration. As tranceiver, it uses the Semtech SX1257 chipset. It is further able to report the measured parameters of the channel from the previous acknowledgement reception from the gateway within the following uplink message. Additionally, it provides the protocol metadata, such as \gls{phy} and \gls{mac} header fields (e.g. the uplink and downlink counters) in order to detect out-of-order receptions or erroneous messages.
As gateway, we used the Wirnet iFemtoCell from Kerlink, which is equipped with the Semtech SX1301 \gls{lora} demodulator for parallel reception of \gls{lora} streams and the Semtech SX1257 transceiver chipset which is mainly used for transmission of downlink control messages.\\
Our measurements are conducted with respect to indoor scenarios, emulating applications such as e.g. smart metering, where end devices are placed within basement and the gateway is positioned several floors above (here fourth floor of the respective building).
The data packets received by the gateway are then forwarded to the network server and application server (\gls{mqtt} broker) respectively, from where they are finally acquired using an \gls{mqtt} client. In total, there were $22912$ samples recorded (equalling $179$ key candidates of length $128$ Bit), and the end devices were configured to transmit test messages at a periodicity of $10$ seconds. We conducted our measurements within the EU868 ISM band and used a channel bandwidth of $125$ kHz. The spreading factor was set to value SF8 (spreading sequence of length $2^8$). Fig. \ref{rssi_captures} shows the log of \gls{rssi} samples captured by both sides, the gateway (\gls{rssi} UE) and end device (\gls{rssi} GW). It is to note, that there is a constant gain within the measurements at the gateway side (see Fig. \ref{rssi_absolute}) resulting from the advanced receiver hardware compared to that of the UE side. Therefore, we pre-process the data samples on both sides, gateway and end device, by removing mean and transform them to unit variance (see Fig. \ref{rssi_scaled}). This step is done on each of the $128$ sample blocks, from which the key bits will be derived respectively.

\begin{table}[h!]
\label{table_results}
\caption{Average \gls{bdr} results}
 \begin{center}
 \begin{tabular}{||c||c||}
 \hline
 Quantization scheme & \gls{bdr}  \\ [0.5ex] 
 \hline\hline
 Threshold & 0.1822 \\ 
 \hline
 Difference & 0.1165 \\
 \hline
 \end{tabular}
 \end{center}
\end{table}
\subsection{Results \& Discussion}
After pre-processing the raw \gls{rssi} samples and splitting them into blocks of length $128$, the quantization step follows, where the values are transformed into key bits. Here, we apply two different quantization schemes, one based on the threshold method, where \gls{rssi} samples above the block threshold yield key bits as $1$, and sample values below the threshold yield key bit $0$. The second method yields a key bit as $1$, if the \gls{rssi} value is larger compared to the previous value, and key bit $0$ vice versa. Therefore it is referred to as difference method. Then the \gls{bdr} and \gls{kgr} are calculated respectively for each block. Table I shows the results in terms of the \gls{bdr} before the information reconciliation stage depending on the quantization method. It can be noted, that for both methods the \gls{bdr} stays below the limit of $20\%$, for which information reconciliation is still possible. The difference method however yields a \gls{bdr} of $11.65\%$, which is $63.9\%$ lower compared to the threshold method at $18.22\%$ \gls{bdr}. It has also to be noted, that no further pre-processing in order to enhance the channel reciprocity was applied, and therefore the \gls{bdr} might still be improved by the utilization of such schemes. Fig. \ref{kgr_dep_bdr} shows the respective \gls{kgr}, again for both methods and depending on the upper limit with respect to the preliminary key \gls{bdr}. In that case for a maximum \gls{bdr} of $5\%$, only the difference method yields adequate key candidates at a \gls{kgr} of $0.0088$ bit/s which results in a total duration of $\approx 4$ hours on average. However, at a tolerable \gls{bdr} of $10\%$, the \gls{kgr} of $0.0534$ bit/s already reflects to a key generation duration of $\approx 40$ minutes on average for that quantization method. If the maximum of tolerable preliminary key \gls{bdr} is allowed, where it is still possible to recover the key bits by information reconciliation ($20\%$), the key generation times result in $31.5$ minutes for the threshold method at a \gls{kgr} of $0.0677$ bit/s, and to $25.15$ minutes for the difference method at a \gls{kgr} of $0.0848$ bit/s. 

\section{Conclusion}
\label{conclusion}
Within this work, we proposed the application of \gls{physec} based key management within \gls{lorawan}. The experimental results show, that it can be a promising approach in order to enable key management at a low cost in terms of energy consumption and complexity, compared to conventional key management solutions. There is even still some potential to further improve the performance in terms of \gls{bdr} by using adequate pre-processing such as reciprocity enhancement strategies. Further, it is necessary to investigate additional scenarios, such as e.g. outdoor deployments.The information reconciliation stage can be improved, by e.g. designing codes with as less parity bit exchange as possible in order to reduce the risk of an attacker gaining key knowledge, as well as reduction in energy consumption for the additional messaging overhead. 


\bibliographystyle{ieeetr}
\bibliography{ngna_ref}

\begin{thebibliography}{10}

\bibitem{NIST2001}
NIST, ``{Advanced Encryption Standard (AES)},'' 2001.

\bibitem{lora_spec}
``Lorawan™ 1.1 specification,'' 2017.

\bibitem{IETF2006}
IETF, {\em RFC 4493, The AES-CMAC Algorithm}, 2006.

\bibitem{Zhang2016}
J.~Zhang, R.~Woods, T.~Q. Duong, A.~Marshall, Y.~Ding, Y.~Huang, and Q.~Xu,
  ``Experimental study on key generation for physical layer security in
  wireless communications,'' in {\em IEEE Access}, vol.~4, pp.~4464--4477,
  2016.

\bibitem{Yang2015}
N.~Yang, L.~Wang, G.~Geraci, M.~Elkashlan, J.~Yuan, and M.~D. Renzo,
  ``Safeguarding 5g wireless communication networks using physical layer
  security,'' in {\em IEEE Communications Magazine}, vol.~53, pp.~20--27, 2015.

\bibitem{Lipps2019}
C.~Lipps, M.~Strufe, S.~B. Mallikarjun, and H.~D. Schotten, ``Physec in
  cellular networks: Enhancing security in the iiot,'' in {\em ECCWS 2019 18th
  European Conference on Cyber Warfare and Security}, 2019.

\bibitem{Wang2018}
L.~Wang, J.~Liu, M.~Chen, G.~Gui, and H.~Sari, ``Optimization-based access
  assignment scheme for physical-layer security in d2d communications
  underlaying a cellular network,'' in {\em IEEE Transactions on Vehicular
  Technology}, vol.~67, pp.~5766--5777, 2018.

\bibitem{Geraci2014}
G.~Geraci, H.~S. Dhillon, J.~G. Andrews, J.~Yuan, and I.~B. Collings,
  ``Physical layer security in downlink multi-antenna cellular networks,'' in
  {\em IEEE Transactions on Communications}, vol.~62, pp.~2006--2021, 2014.

\bibitem{Guillaume2015}
R.~Guillaume, F.~Winzer, A.~Czylwik, C.~T. Zenger, and C.~Paar, ``Bringing
  phy-based key generation into the field: An evaluation for practical
  scenarios,'' in {\em IEEE 82nd Vehicular Technology Conference (VTC Fall)},
  2015.

\bibitem{Ambekar2012}
A.~Ambekar, N.~Kuruvatti, and H.~D. Schotten, ``Improved method of secret key
  generation based on variations in wireless channel,'' in {\em IEEE
  International Conference on Systems, Signals and Image Processing (IWSSIP),
  Vienna, Austria}, 2012.

\bibitem{Zenger2014}
C.~T. Zenger, M.-J. Chur, J.-F. Posielek, C.~Paar, and G.~Wunder, ``A novel key
  generating architecture for wireless low-resource devices,'' in {\em
  In­ter­na­tio­nal Work­shop on Se­cu­re In­ter­net of Things
  (SIoT)}, 2014.

\bibitem{Zenger2015a}
C.~T. Zenger, J.~Zimmer, M.~Pietersz, J.-F. Posielek, and C.~Paar, ``Exploiting
  the physical environment for securing the internet of things,'' in {\em New
  Se­cu­ri­ty Pa­ra­digms Work­shop (NSPW)}, 2015.

\bibitem{Lipps2018}
C.~Lipps, A.~Weinand, D.~Krummacker, C.~Fischer, and H.~D. Schotten, ``Proof of
  concept for iot device authentication based on sram pufs using atmega
  2560-mcu,'' in {\em 2018 1st International Conference on Data Intelligence
  and Security (ICDIS)}, pp.~36--42, 2018.

\bibitem{Lipps2020}
C.~Lipps, P.~Ahr, and H.~D. Schotten, ``How to secure the communication and
  authentication in the iiot: A sram-based hybrid cryptosystem,'' in {\em
  European Conference on Cyber Warfare and Security (ECCWS)}, June 2020.

\bibitem{AdefemiAlimi2020}
K.~O. Adefemi~Alimi, K.~Ouahada, A.~M. Abu-Mahfouz, and S.~Rimer, ``A survey on
  the security of low power wide area networks: Threats, challenges, and
  potential solutions.,'' in {\em Sensors}, vol.~20, Oct. 2020.

\bibitem{Robyns2017}
P.~Robyns, E.~Marin, W.~Lamotte, P.~Quax, D.~Singelée, and B.~Preneel,
  ``Physical-layer fingerprinting of lora devices using supervised and
  zero-shot learning,'' in {\em 10th ACM Conference on Security and Privacy in
  Wireless and Mobile Networks}, 2017.

\bibitem{Ruotsalainen2020}
H.~Ruotsalainen, J.~Zhang, and S.~Grebeniuk, ``Experimental investigation on
  wireless key generation for low-power wide-area networks,'' in {\em IEEE
  Internet of Things Journal}, vol.~7, pp.~1745--1755, 2020.

\bibitem{Zhang2018}
J.~Zhang, A.~Marshall, and L.~Hanzo, ``Channel-envelope differencing eliminates
  secret key correlation: Lora-based key generation in low power wide area
  networks,'' in {\em IEEE Transactions on Vehicular Technology}, vol.~67,
  pp.~12462--12466, 2018.

\bibitem{Xu2019a}
W.~Xu, S.~Jha, and W.~Hu, ``Lora-key: Secure key generation system for
  lora-based network,'' in {\em IEEE Internet of Things Journal}, vol.~6,
  pp.~6404--6416, 2019.

\bibitem{Seller2016}
O.~B.~A. Seller, ``Wireless communication method,'' Patent US20160094269A1,
  March 2016.

\bibitem{lora_regional}
``Lorawan™ 1.1 regional parameters,'' 2017.

\end{thebibliography}

\end{document}